# Application of Secondary Neutral Mass Spectrometry in the investigation of doped perovskites

K. Vad[1*], J. Hakl[1], A. Csik[1], S. Mészáros[1], M. Kis-Varga[1], G.A. Langer[2], Á. Pallinger[3], M. Bódog[4]

[1]*Institute of Nuclear Research, Hungarian Academy of Sciences, H-4001, Debrecen, P.O. Box 51, Hungary*

[2]*Department of Solid State Physics, University of Debrecen, H-4010, Debrecen P.O. Box 2, Hungary*

[3]*Research Institute for Solid State Physics and Optics of the Hungarian Academy of Sciences, H-1525, Budapest, P.O. Box 49, Hungary*

[4]*Pannon University, H-8201, Veszprém, P.O. Box 158, Hungary*

**Abstract**

Electromagnetic properties of doped perovskites depend sensitively on the doping level. Both the superconducting transition temperature of $Bi_2Sr_2Ca(Pr)Cu_2O_{8+\delta}$ compounds and the magnetic and electronic transport properties of $La(Sr)Co(Fe)O_3$ perovskites change dramatically with the doping level. Apart from doping, oxygen deficiency is influenced by the details of preparation processes such as calcination and sintering. Simultaneous determination of constituents is of crucial importance from sample characterization point of view. Quantitative analysis of perovskite oxides can be performed by Secondary Neutral Mass Spectrometry (SNMS) which is a suitable technique to measure the chemical composition of almost any sample because the flux of atoms sputtered from the sample is representative of the stoichiometry of the top-most layers. The composition and oxygen content of $Bi_2Sr_2Ca_{0.86}Pr_{0.14}Cu_2O_{8.4}$ and $La_{1-x}Sr_xCo_{0.975}Fe_{0.025}O_{3-\delta}$, where $0 \leq x \leq 0.5$, were determined by

[*] Corresponding author: *Fax*: +36-52-416181
*E-mail address*: vad@atomki.hu (K. Vad)

SNMS. The results show that the method is equally applicable for insulating and conducting compounds. The observed electromagnetic properties reflect well the compositions obtained experimentally.

**1. Introduction**

Doped perovskites have become a highlight of modern solid state physics due to the discovery of high-temperature superconductivity (HTS) and colossal magnetoresistance (CMR). Some types of perovskites (cuprates) show superconductivity, others (manganates, cobaltates) show CMR. The common feature of these two families of compounds is that their electromagnetic properties depend sensitively on doping level. HTS cuprates are layered perovskite oxides. The basic structure of all high-temperature superconductors consists of three types of layers: charge reservoir layers, superconducting layers and mediating layers [1]. In the parent $Bi_2Sr_2CaCu_2O_8$ compound these are the Bi-Sr-O layer, the Cu-O layer and the Ca layer, respectively. Superconductivity develops exclusively in the two Cu-O planes separated from each other by the mediating Ca layer, and the superconducting critical temperature ($T_c$) depends on the distance between these Cu-O layers. Hall measurements show that the charge carriers in $Bi_2Sr_2CaCu_2O_8$ are holes, and show also that some doping process is needed to create them. Hole doping in the Cu-O planes can be achieved by increasing the oxygen content of the sample. The parent compound $Bi_2Sr_2CaCu_2O_8$ is not superconducting itself, it becomes only superconducting with subsequent oxygen doping in the Bi-O layer. About 0.2 excess oxygen per formula unit, i.e. $Bi_2Sr_2CaCu_2O_{8+\delta}$ with $\delta \sim 0.2$, is necessary to enter the superconducting phase. After oxygen doping, using a suitable nonisovalent substitution in the parent compound, the hole-carrier concentration and the degree of the concomitant charge transfer to $CuO_2$ layers can be subtly modified. For example, substitution of divalent Ca by trivalent Pr depresses superconductivity in $Bi_2Sr_2CaCu_2O_{8+\delta}$ since the higher valence Pr will contribute electrons to $CuO_2$ planes and fill the mobile holes responsible for conduction. The effect of two Pr atoms counterbalances the effect of one oxygen atom.

Transition-metal oxides with composition $ABO_3$ make up the other perovskite family. The compound $LaCoO_3$ is a typical member of this family. It is of particular interest because it shows a phase transition from insulating state to ferromagnetic metallic phase as the charge carrier doping level increases [2]. Substituting divalent Sr for trivalent La amounts to hole doping in $LaCoO_3$. The excess positive charge can be absorbed by the oxidation of Co, i.e. forming of $Co^{4+}$ cations instead of $Co^{3+}$, and by the emergence of oxygen vacancies. The exact oxygen content of nominal $La_{1-x}Sr_xCoO_3$ samples is difficult to determine, but they can be safely taken to be mixed-valent oxides. Double-exchange mechanism has been invoked to explain the simultaneous insulator-to-metal and paramagnetic-to-ferromagnetic transitions at $x = 0.18$. Magnetoresistance has been of paramount importance in most of the work on perovskites in the past decade. The 'colossal' effects in manganates were observed [3,4] in high magnetic fields, far below room temperature. $La_{1-x}Sr_xCoO_3$ was also found to show colossal magnetoresistance.

The electromagnetic properties of doped perovskites - the superconducting and other phase transition temperatures, the temperature and magnetic field dependence of the resistivity, etc. - depend sensitively on the doping level and oxygen content of the sample. The interest in analytical techniques for accurate determination the elemental concentration of perovskite oxides increases with improvements in experimental technique. Secondary Neutral Mass Spectrometry (SNMS) is a successful technique for the analysis of perovskites. Its high sensitivity makes it suited for the elemental characterization of metallic and insulating samples. Such characterization is indispensable for the understanding of doping processes in perovskites. Special attention is to be paid to the oxygen content which can vary at fixed doping level.

## 2. Experimental details

The samples studied were prepared by the conventional powder-metallurgy approach, i.e. mixing and heating the constituent oxides and carbonates. The details of preparation of Pr doped $Bi_2Sr_2Ca(Pr)Cu_2O_{8+\delta}$ compound were described in Refs. [5,6]. It was relatively easy to produce cleaved samples from the batch prepared, a consequence of its extreme quasi-two-dimensionality. Unfortunately, the two-dimensional nature of the material also resulted in relatively thin single crystals being grown; typical thicknesses were of a few μm in the *c* direction. This precluded or made very difficult to study of the $Bi_2Sr_2CaCu_2O_{8+\delta}$ family by experimental techniques which required a large volume of material. Our measurements were performed on a crystal of size 370x400x5 μm$^3$. This sample was cut from a larger one which had previously been prepared for low temperature, high magnetic field transport and magnetization measurements.

The bulk polycrystalline samples of cobaltates with nominal composition $La_{1-x}Sr_xFe_{0.025}Co_{0.975}O_{3-\delta}$ ($x$ = 0, 0.15, 0.18, 0.2, 0.22, 0.25) were prepared by standard ceramic method as described in Ref. [7]. A stoichiometric mixture of $La_2O_3$, $Co_3O_4$, $SrCO_3$ and Fe powder (dissolved in nitric acid) was homogenized and pressed into pellets of 10 mm diameter under a pressure of 500 MPa, then calcined in air at 980 $^o$C for 168 hours. After cooling (160 $^o$C/h) the samples were ground again and pressed into pellets. At last a second annealing was applied at 1220 $^o$C for 2 hours. At the end of the procedure the pellets were cooled down by a slow cooling rate of 50 °C/h. Powder x-ray diffraction data confirmed the quality of the samples.

The stoichoimetric ratio of the samples was determined by secondary neutral mass spectrometry. SNMS is a multielemental and simultaneous technique based on the measurement of secondary neutral elements sputtered from the surface of the sample. The SNMS is a destructive analysis technique. The sample surface is sputtered by an ion beam. Most of the sputtered material consists of neutral atoms. These neutral atoms ejected from the surface are detected by post-ionization using Electron Cyclotron Wave Resonance (ECWR) plasma. This procedure reduces the matrix effect and allows a much better quantitative estimate of the stoichiometry than Secondary Ion Mass Spectrometry. It is a suitable technique to measure the chemical composition of almost any sample, because the flux of atoms sputtered from the sample is representative of the stoichiometry. This is in contrast to X-ray Photoelectron Spectroscopy (XPS) systems combined with an ion gun, where preferential sputtering makes analysis more difficult. A further advantage of SNMS is its applicability to insulators, which is lacking in other methods based on electron emission or excitation by a primary beam of electrons (such as Auger Electron Spectroscopy or XPS).

SNMS is a very rapid method compared with other surface analysis techniques and does not suffer from many of the problems which exist in other methods. It can be used for analysing small samples with a high depth resolution. In our experiments the SNMS measurements were performed by an instrument, type INA-X, which was developed by SPECS GmbH, Berlin.

## 3. Results and discussion

Fig. 1 shows the mass spectrum of $Bi_2Sr_2Ca_{0.86}Pr_{0.14}Cu_2O_{8.4}$ sample. The most exciting challenge in measurements of $Bi_2Sr_2Ca(Pr)Cu_2O_{8+\delta}$ samples was the small mass. Because the sample dimensions were a few hundred μm and the thickness was only a few μm, the sample mass was 1 μg. The picture of this sample is shown in the inset of Fig. 1. The inner circular part was sputtered with Ar ions extracted from the plasma. The bombarding energy was 600V. The chemical composition was determined using relative sensitivity factors of constituents [8]. The mass peaks in Fig. 1 can be divided into three groups. The first group originated from the vacuum chamber and from the organic solvent. The second group is comprised of the peaks characterizing the metal constituents. The third one has the oxide peaks. It is evident from Figure 1, that the intensities of oxide molecules are not negligible. Moreover, as we checked before starting measurements on a Ta sample, oxygen leaves the sample exclusively in the form of oxides and not as O atoms or $O_2$ molecules. So, the oxygen content of the sample was determined by evaluating the oxides peaks. According to our measurements we found $Bi_2Sr_2Ca_{0.86}Pr_{0.14}Cu_2O_{8.4}$ as the formula unit of the HTS compounds.

Likewise, oxygen leaves the cobaltate samples during sputtering in the form of molecular clusters. Of the peaks due to La and LaO, the intensity of LaO was 3 times higher. From the mass spectra measured by SNMS the correct elemental concentrations were determined as shown in Fig. 2. It is clear from the measurements that the oxygen content depends on the doping level of Sr. This result was checked by Electron Dispersive X-Ray (EDX) analysis, which also supported the Sr dependence of oxygen content.

## 4. Conclusion

The electromagnetic properties of perovskites depend sensitively on doping level. The determination of doping levels with the required accuracy calls for a precise composition analysis. We demonstrated that Secondary Neutral Mass Spectrometry is an adequate technique for the analysis of perovskites, and that SNMS can be employed for the chemical analysis of very tiny samples in both metallic and insulating phases. To determine the oxygen content, which is the crucial characteristic of perovskites, the escaped oxide molecules have to be taken into account.

**Acknowledgement**

The authors are very grateful to L. Forró (EPF, Lausanne) for preparing the doped HTS sample. This work was partially supported by the grant of Hungarian Scientific Research Fund (Grant No. K-62866).

**Figure captions**

Figure 1. Mass spectrum of $Bi_2Sr_2Ca_{0.86}Pr_{0.14}Cu_2O_{8.4}$. The inset shows the sample with the circular shaped area on it surface where the analysis was made.

Figure 2. The analysis of $La_{1-x}Sr_xCo_{0.975}Fe_{0.025}O_{3-\delta}$ samples by SNMS. The oxygen content of the samples also was checked by Electron Dispersive X-ray analysis (EDX) as it is shown. The experimental error in the measurements was approximately ± 3 %.

Figure 1. K. Vad, J. Hakl, A. Csik, S. Mészáros, M. Kis-Varga, G.A. Langer, Á. Pallinger, M. Bódog

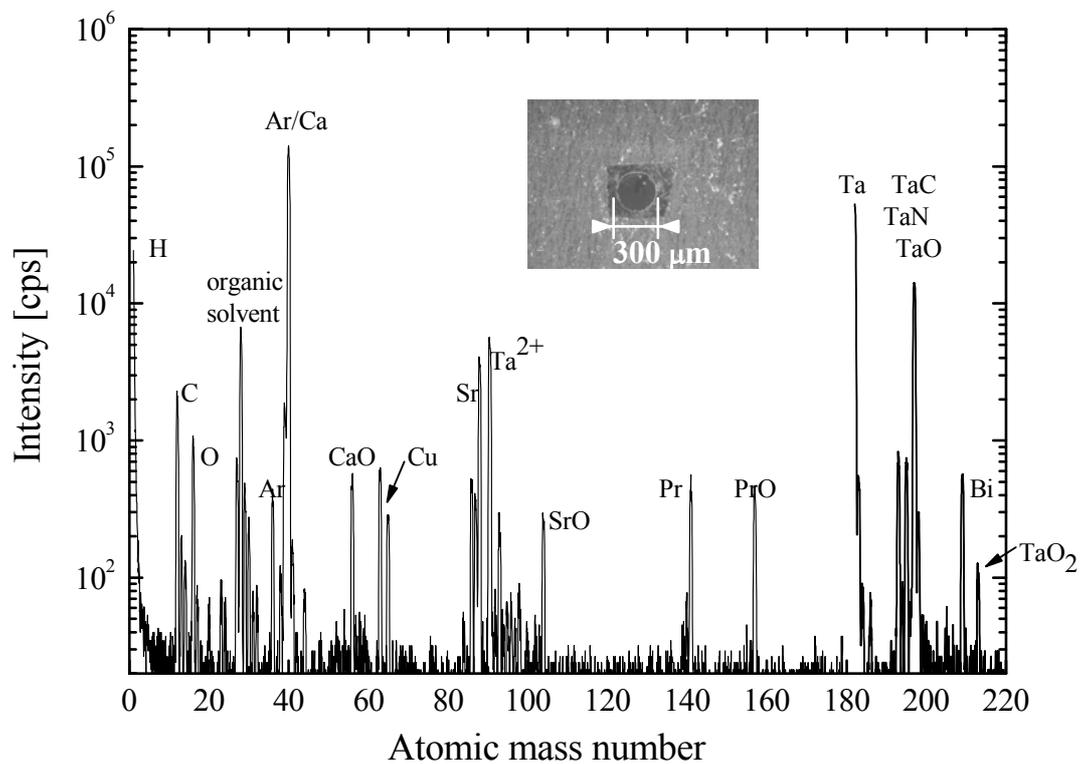

Figure 2. K. Vad, J. Hakl, A. Csik, S. Mészáros, M. Kis-Varga, G.A. Langer, Á. Pallinger, M. Bódog

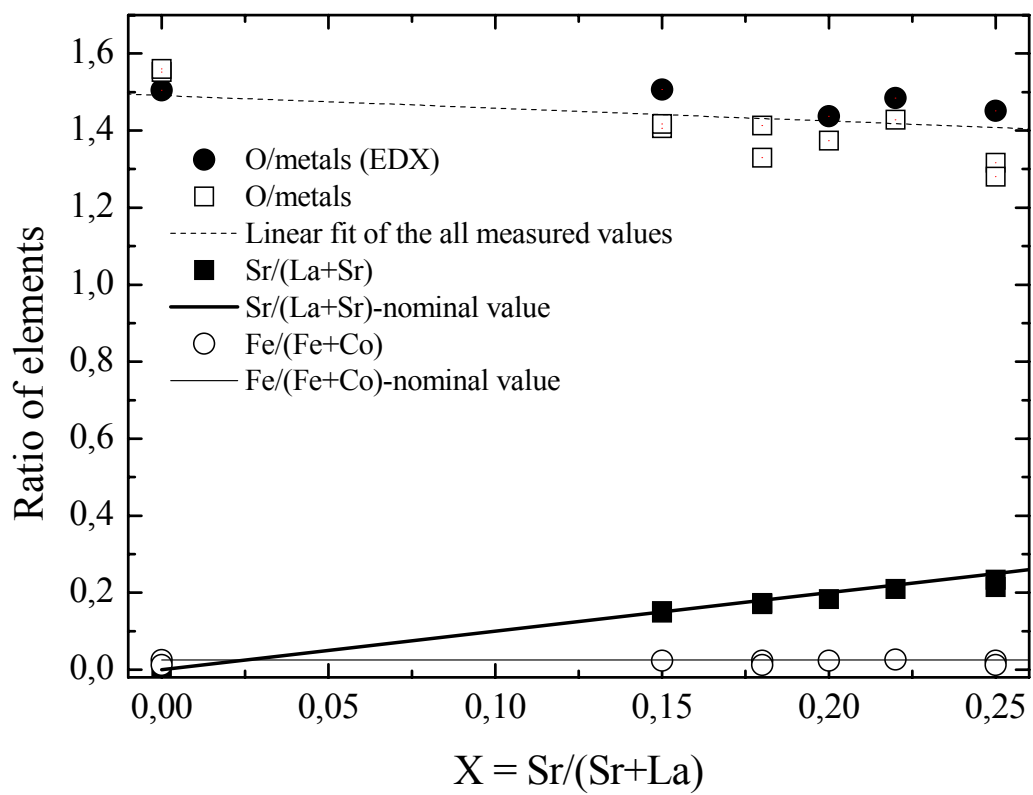